\documentclass[aps, prl, twocolumn, floatfix, superscriptaddress, showpacs, amssymb, reprint]{revtex4-1}

\usepackage{graphicx}
\usepackage{color}

\begin{document}

\title{Investigation of unconventional reconstruction and electronic properties on the Na$_{2}$IrO$_{3}$ surface}
\author{F. L\"upke}
\affiliation{IV. Physikalisches Institut, Georg-August-Universit\"at G\"ottingen, D-37077 G\"ottingen, Germany}
\affiliation{Peter Gr\"unberg Institut (PGI-3), Forschungszentrum J\"ulich, 52425 J\"ulich, Germany}
\author{S. Manni}
\affiliation{I. Physikalisches Institut, Georg-August-Universit\"at G\"ottingen, D-37077 G\"ottingen, Germany}
\affiliation{Experimental Physics VI, Center for Electronic Correlations and Magnetism, University of Augsburg, D-86159 Augsburg, Germany}
\author{S. C. Erwin}
\affiliation{Center for Computational Materials Science, Naval Research Laboratory, Washington, DC 20375, USA}
\author{I. I.  Mazin}
\affiliation{Center for Computational Materials Science, Naval Research Laboratory, Washington, DC 20375, USA}
\author{P. Gegenwart}
\affiliation{Experimental Physics VI, Center for Electronic Correlations and Magnetism, University of Augsburg, D-86159 Augsburg, Germany}
\author{M. Wenderoth}
\altaffiliation[Corresponding author.\\]{\urlstyle{same}\url{mwender@gwdg.de}}
\affiliation{IV. Physikalisches Institut, Georg-August-Universit\"at G\"ottingen, D-37077 G\"ottingen, Germany}

\date{\today}


\begin{abstract}
  Na$_{2}$IrO$_{3}$ is an intriguing material for which spin-orbit
  coupling plays a key role. Theoretical predictions, so far
  unverified, have been made that the surface of Na$_{2}$IrO$_{3}$
  should exhibit a clear signature of the quantum spin Hall effect. We
  studied the surface of Na$_{2}$IrO$_{3}$  using
  scanning tunneling microscopy and
  density-functional theory calculations. We observed atomic level resolution of the surface and  two types of
  terminations with different surface periodicity and Na content. By
  comparing bias-dependent experimental topographic images to
  simulated images, we determined the detailed atomistic structure of both
  observed surfaces. One of these reveals a strong relaxation to the
  surface of Na atoms from the subsurface region two atomic layers
  below.  Such dramatic structural changes at the surface cast doubt
  on any prediction of surface properties based on bulk electronic
  structure. Indeed, using spatially resolved tunneling spectroscopy
  we found no indication of the predicted quantum spin Hall behavior.
\end{abstract}

\maketitle

Novel states with unusual topological and frustrated properties have
recently been predicted to arise in heavy transition-metal oxides,
such as iridates, from a combination of interactions---spin-orbit
coupling, Coulomb correlations, Hund's rule coupling, and one-electron
hopping---with comparable energy scales
\cite{Kim,Shitade,Pesin,Jackeli1,Jiang,Mazin12, Foyev13}.
Na$_2$IrO$_3$ is a prototypical material in the iridate family. It
consists of an alternating stacking of honeycomb Ir$_{2}$NaO$_{6}$
layers separated by hexagonal Na$_{3}$ layers \cite{Singh10}. While
many works have concentrated on unusual magnetic properties of the
bulk material, the surface of Na$_{2}$IrO$_{3}$ may also reveal
unusual physics. For example, recent theoretical work predicts quantum
spin Hall (QSH) behavior in Na$_{2}$IrO$_{3}$ \cite{Shitade}.  The
resulting band topology of the bulk should lead to helical edge states
at the surface, which would be manifested experimentally by the
closing of the band gap \cite{Shitade}.

The prediction of QSH behavior was based on a tight-binding model
derived from the bulk electronic structure \cite{Shitade}. It was
subsequently shown that the bulk states depend very sensitively on the
assumed geometry---in particular, on the positions of Na and
corresponding rotations of the IrO$_{6}$ octahedra \cite{Foyev13}.  An
important question, so far unaddressed, is whether the geometry at the
surface of Na$_2$IrO$_3$ is sufficiently similar to the bulk to support the assumptions
underlying the QSH prediction.  In this Letter, we used scanning
tunneling microscopy (STM) and spectroscopy (STS) together with
density-functional theory (DFT) calculations to address this question.
We also tested the QSH prediction directly by spectroscopically
probing for gap closure on the surface.
Our most important findings are that (1) the surface of Na$_2$IrO$_3$ strongly reconstructs in
a highly unconventional manner, which very likely undermines the
conditions for QSH behavior; (2) the surface band gap does {\em not}
close, which establishes that QSH behavior is indeed not realized at the
Na$_2$IrO$_3$ surface.

\begin{figure}
\includegraphics[width=7cm]{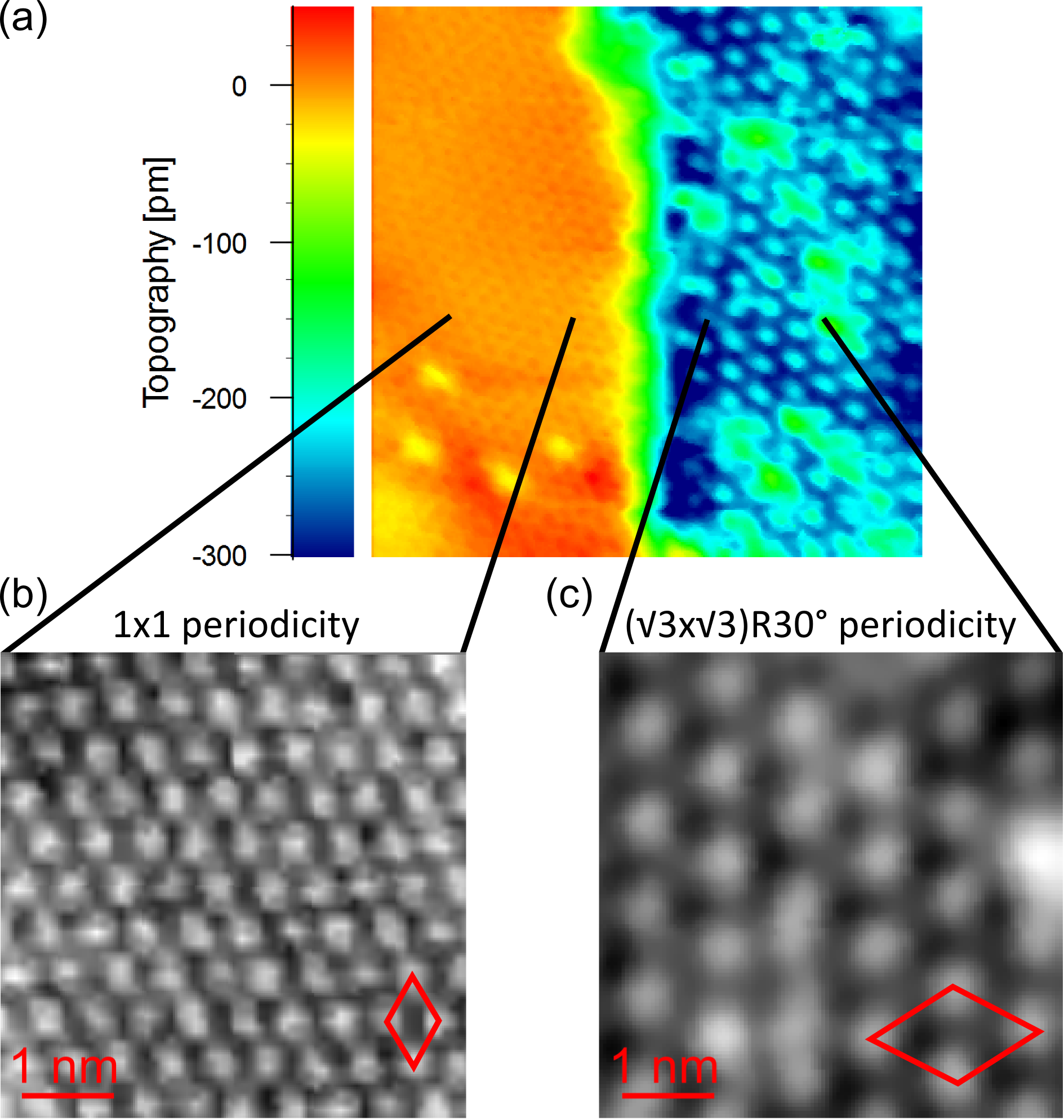}  
\caption{(Color online) Constant-current STM topographic images at $V_{%
\mathrm{bias}}=+2$ V. (a) Two different surface configurations with well-defined
boundaries are observed. (b) Detail of the $1\times 1$ surface and (c) $(%
\protect\sqrt{3}\times \protect\sqrt{3})\mathrm{R30^{\circ }}$ surface, with
the respective unit cells indicated. The $1\times 1$ surface exhibits
long-range order, while the $(\protect\sqrt{3}\times \protect\sqrt{3}%
)\mathrm{R30^{\circ }}$ surface exhibits many defects and 
only local order.}
\label{1}
\end{figure}

\begin{figure*}
\includegraphics[width=16cm]{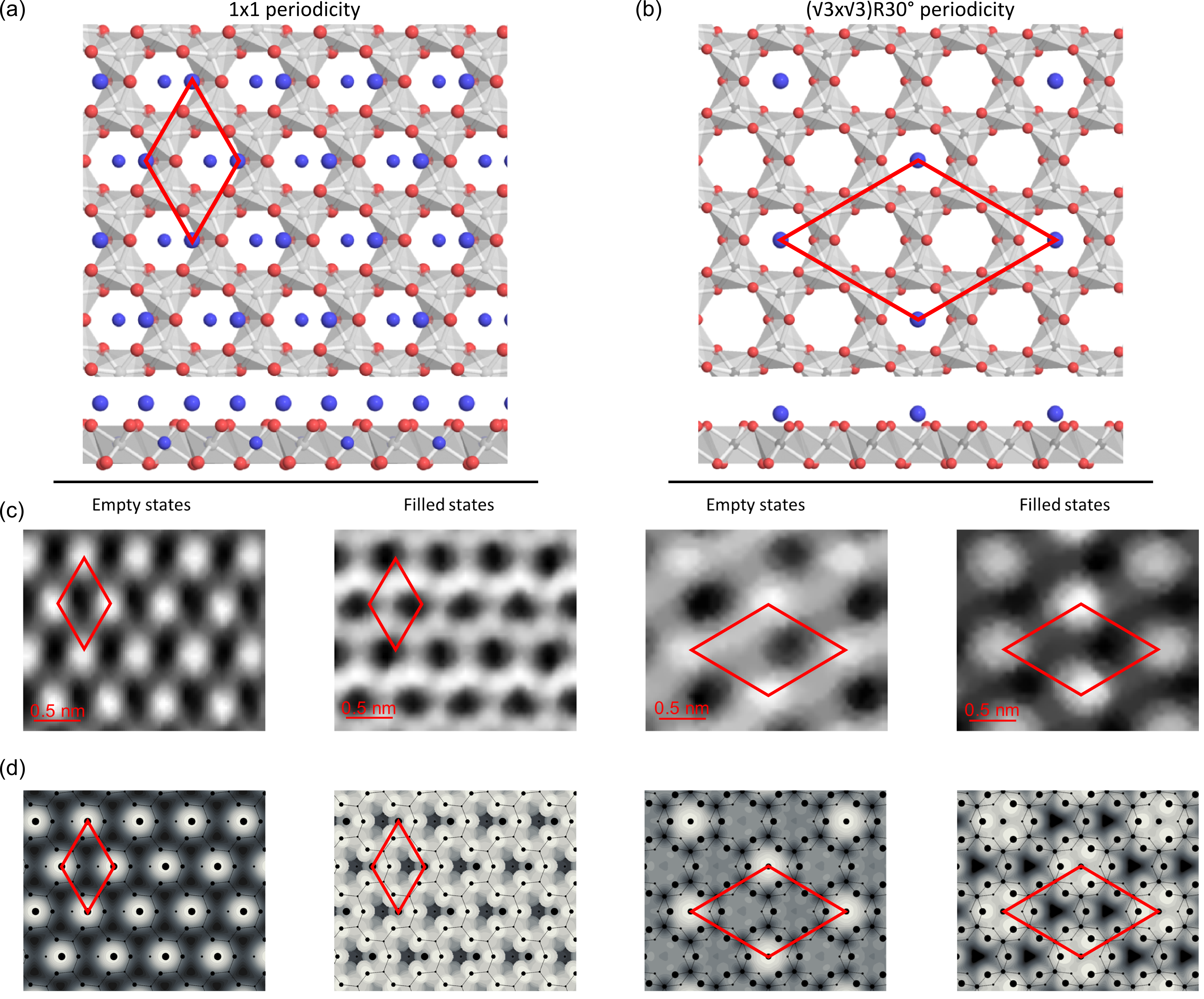}  
\caption{(Color online) (a,b) Theoretically determined structural
  models for $1\times 1$ and $(\protect\sqrt{3}\times
  \protect\sqrt{3})\mathrm{R30^{\circ }}$ surfaces viewed from the top and side, respectively (Na: blue, Ir:
  gray, O: red). Na atoms in the topmost layer are shown larger for
  clarity. (c) Experimental and (d) theoretically simulated STM images
  at $V_{\mathrm{bias}}=\pm 2$ V. On the $1\times 1$ surface, there is
  a clear contrast reversal between empty and filled states. On the
  $(\protect\sqrt{3}\times \protect\sqrt{3})\mathrm{R30^{\circ }}$
  surface, the dominant bright features for empty and filled images
  are located at the same positions.}
\label{2}
\end{figure*}

Na$_{2}$IrO$_{3}$ surfaces were prepared by in situ cleaving at
a base pressure of $p<10^{-10}$ mbar. After immediate transfer to a
home-built low temperature scanning tunneling microscope (STM)
operating at 80 K, the surface was investigated by STM and STS. The
latter allows mapping of occupied and unoccupied states as well as
simultaneous access to spatial variations in the electronic surface
properties \cite{Feenstra1, Ye}. On the freshly cleaved surface, two
different stable surface terminations were found in constant-current
topographic (CCT) measurements both showing atomic level resolution (Fig.~\ref{1}). The periodicity of the terminated surfaces suggests cleaving along the $ab$-plane of the crystal. One termination shows the
periodicity of bulk Na$_{2}$IrO$_{3}$ and is labeled as $1\times
1$ in the following. The second termination shows a $(\sqrt{3}\times\sqrt{3})\mathrm{R30^{\circ }}$ reconstruction with respect to the
$1\times 1$.  The two terminations occur in roughly the same
proportion.  While the $1\times 1$ surface shows a long-range
periodic structure over tens of nanometers, the $(\sqrt{3}\times
\sqrt{3})\mathrm{R30^{\circ }}$ surface exhibits lots of defects and
is well-ordered only on the scale of a few nanometers.

The cleavage process almost certainly leaves the strongly bonded
IrO$_6$ octahedra intact. Hence, the observation of two different
surface terminations suggests that cleaving creates two metastable
surfaces with different Na coverage and hence different periodicity.
We used DFT to determine the equilibrium geometries of different
candidate surfaces with varying Na content in the top layer as well as
subsurface layers.  Total energies and forces were calculated within
the PBE generalized-gradient approximation using
projector-augmented-wave potentials, as implemented in {\sc vasp}
\cite{kresse_phys_rev_b_1993a,Kresse1996}.  After relaxation, we
simulated STM images using the method of Tersoff and Hamann
\cite{Tersoff}, by integrating the local density of states (LDOS)
within $\pm$2 eV around the Fermi level. The surface of constant
integrated LDOS then corresponds to the ideal STM topography at that
bias voltage.

Figure \ref{2} shows the theoretical equilibrium structure and STM
imagery for the two models that best reproduce the imagery of the two
observed surfaces. The 1$\times$1 surface shown in Fig.\ 2(a) was
constructed by starting from the bulk crystal, which consists of
stacked atomic layers in the stoichiometric sequence
$\cdots|{\rm Na}_3|{\rm O}_3|{\rm Na}_1,{\rm Ir}_2|{\rm O}_3|\cdots$. By cleaving this crystal within the pure Na layer one obtains surfaces
with different relative Na content. For reference, the ideal
stoichiometric surface has a Na$_{3/2}$ surface layer.
We constructed the 1$\times$1
surface in Fig.\ 2(a) by removing one-third of the Na atoms from this
idealized surface. Hence the model in Fig.\ 2(a) is ${\rm Na}_1|{\rm
  O}_3|{\rm Na}_1,{\rm Ir}_2|{\rm O}_3|\cdots$.  The agreement between
simulated and measured constant-current topographies is excellent,
strongly suggesting that this structural model is correct.  In
particular, the strong contrast inversion between empty and filled
states observed experimentally is well reproduced in the simulated
images. Even the detailed topography agrees well: The filled states
appear as a bright honeycomb network, while the empty states appear as
disconnected bright spots. The contrast reversal arises from different
tunneling paths for positive and negative biases, via empty Na states
and filled O states, respectively.

Figure \ref{2}(b) shows an even more substoichiometric surface created
by removing the entire topmost Na layer as well as two-thirds of the Na atoms
in the subsurface Ir-Na layer.
The nominal structure of this surface
is hence  ${\rm O}_3|{\rm
  Na}_{1/3},{\rm Ir}_2|{\rm O}_3|\cdots$, which indeed has the
experimentally observed
$(\sqrt{3}\times\sqrt{3})\mathrm{R30^{\circ }}$ periodicity \cite{footnote1}. Upon
relaxation the Na atoms in the subsurface layer 
move upwards by nearly 2 \AA\ from their ideal
positions. Hence the equilibrium surface structure is actually
 ${\rm Na}_{1/3}|{\rm O}_3|{\rm Ir}_2|{\rm O}_3|\cdots$, as shown in
 Fig.\ 2(b).
 This very large relaxation is confirmed experimentally by
the excellent agreement between experimental and simulated empty-state
images: Both the periodicity and topography are entirely
determined by empty Na states, which only be the case if the Na
atoms have relaxed completely out of the subsurface Ir layer.
The filled-state images are also in very good
agreement. Interestingly, they do not show any contrast reversal. Instead,
the topography in filled-state images actually arises from
oxygen orbitals that are neighbors of surface Na atoms, whereas the
empty-state images are dominated by the unoccupied Na orbitals themselves.

\begin{figure}
\includegraphics[width=7cm]{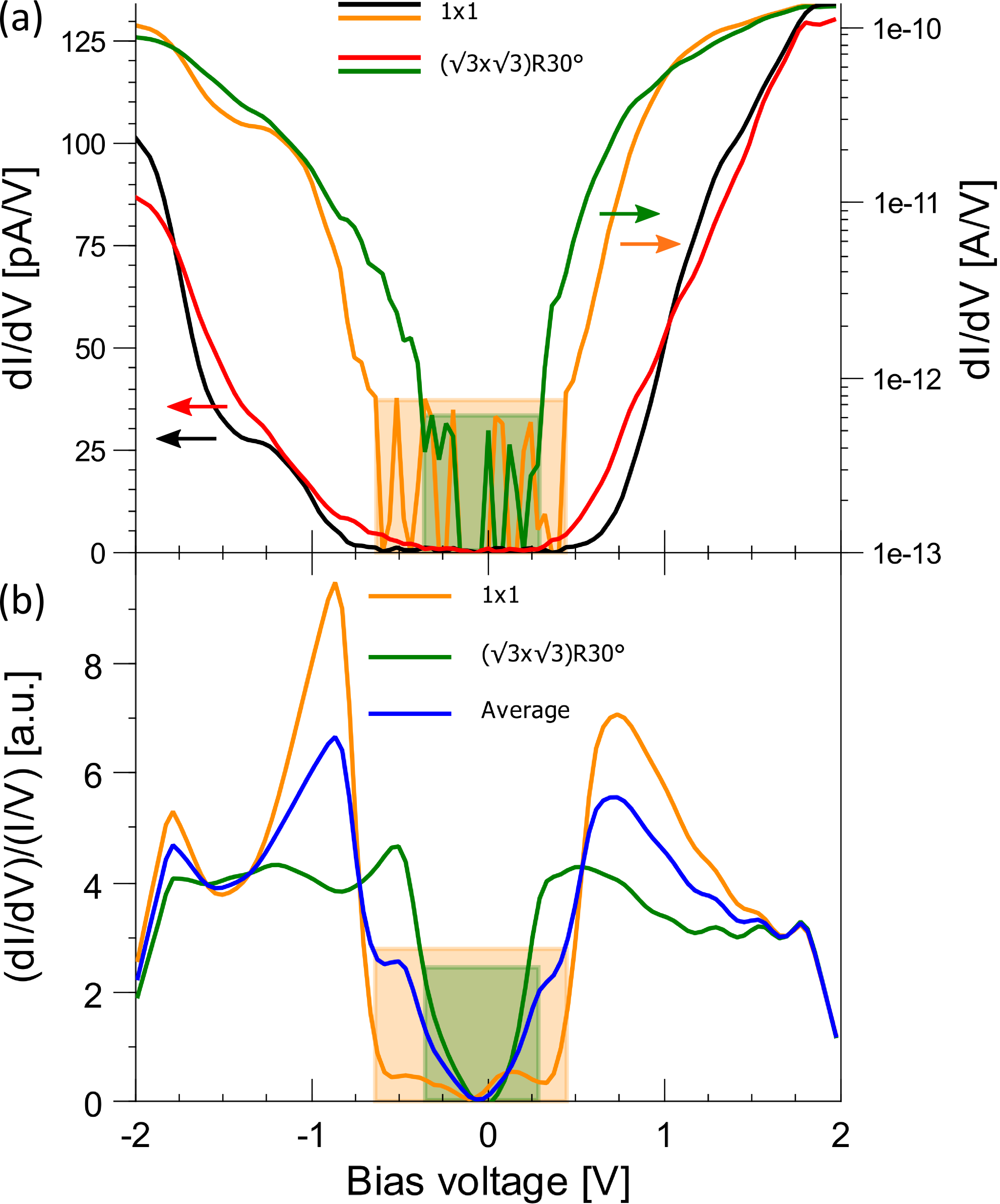}  
\caption{(Color online) (a) Differential conductivity
  spectra of the $1\times 1$ and
  $(\sqrt{3}\times\sqrt{3})\mathrm{R30^{\circ }}$ surface plotted on
  linear and logarithmic scales (left and right, respectively).  The
  region of zero tunnel current within the current detection limit of
  the present setup (700 fA) is identified as the band gap and is indicated in (a) and (b) as shaded areas with colors corresponding to the graphs. The
  $1\times 1$ surface band gap can be estimated to $E_{g}\approx 1$
  eV, while the $(\sqrt{3}\times\sqrt{3})\mathrm{R30^{\circ }}$
  surface shows a smaller band gap of $%
  E_{g}\approx 0.5$ eV. (b) Normalized differential conductivity
  graphs of the two surfaces and their average. Spectral features
  are visible at $V_{\mathrm{bias}}\simeq \pm 1$ V. The averaged spectrum, corresponding to equal surface
  coverages, allows a comparison to ARPES and optical conductivity
  measurements. The data within the shaded areas is below the resolution limit of our setup and apparent features are only the result of the calculation process \cite{Feenstra2}.}
\label{3}
\end{figure}

We next performed a spectral analysis using scanning tunneling spectroscopy (STS) of the two surface terminations to investigate their electronic structure. Figure 3(a)
shows the differential conductivity, $dI/dV,$ of both surface
structures. The current noise level of 700 fA allows us to estimate
the band gaps to be $E_{g}\approx 1$ eV on the $1\times 1$ surface and
$E_{g}\approx 0.5$ eV on the $(\sqrt{3}\times
\sqrt{3})\mathrm{R30^{\circ }}$ surface. One might be tempted to
compare these numbers, especially the less strongly reconstructed
$1\times 1$, with the optical and ARPES band gap of 340 meV reported
in Ref.\ \onlinecite{Comin}. However, this discrepancy must be taken with a
grain of salt because optical absorption is a bulk probe and it is
only natural that the surface gap, after reconstruction and
relaxation, is very different---including the effects of the less
well-screened Hubbard $U$ at the surface.

Regarding ARPES, which is indeed a surface probe, in order to access the excitation gap, in Ref.\ \onlinecite{Comin} the surface of Na$_2$IrO$_3$ was coated with K, to shift the chemical potential into the upper Hubbard band. As we observe a 500 meV difference between different Na$_2$IrO$_3$ surfaces, there is no question that coating with K should have a serious effect on the gap. Moreover, it is also expected that the surface in \cite{Comin} probably also consisted of an ensemble of two reconstructions as found here. Hence to directly compare the surface gap obtained from STS and K-doped ARPES, all these issues need to be considered.

\begin{figure*}
\includegraphics[width=14cm]{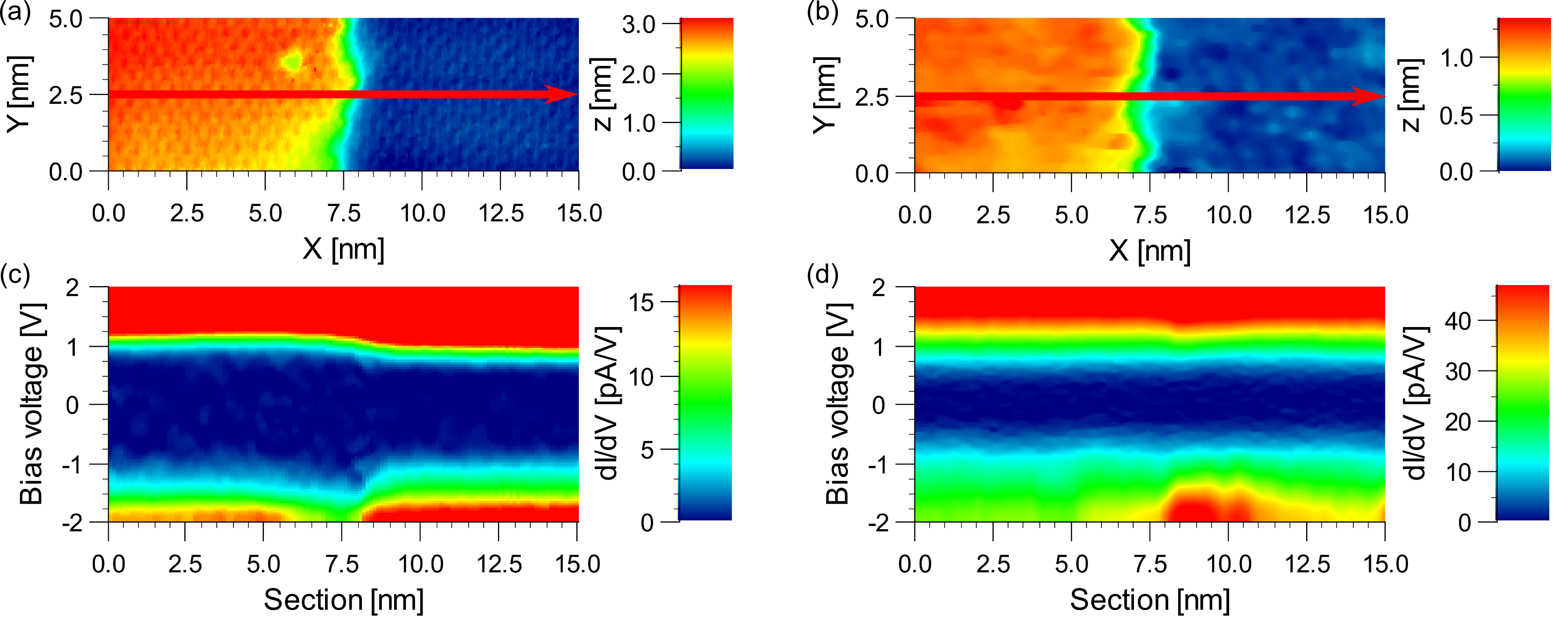}  
\caption{(Color online) (a) STM topography of a surface step between
  two equivalent $1\times 1$ surfaces. The step height is $h\approx
  26.5$ \AA, corresponding to five unit cells of Na$_{2}$IrO$_{3}$.
  (b) Topography of a surface step separating two equivalent
  $(\protect\sqrt{3}\times \protect\sqrt{%
    3})\mathrm{R30^{\circ }}$ surfaces. The step height is $h\approx
  10.6$ \AA, corresponding to two unit cells. (c) Differential
  conductivity spectra along the red line in (a). We observe a smooth
  transition and a slightly larger gap on the upper terrace. The full
  band gap is evident throughout the line scan. (d) Differential
  conductivity spectra along the red line in (b). The band gap is
  maintained across the step edge with only small fluctuations within
  the valence and conduction bands.}
\label{4}
\end{figure*}

Comin \emph{et al.} pointed out that the one-electron hopping, combined with spin-orbit coupling, alters the band structure substantially compared with the spin-orbit-only picture of two $J_{\mathrm{eff}}=1/2$ and $3/2$ level.
In particular, it creates a pseudogap in the center of the $J_{\mathrm{eff}}=1/2$ band, which is then enhanced into a full gap by Hubbard correlations \cite{Comin}. Similarly, the $J_{\mathrm{eff}}=3/2$ quartet is, because of hybridization with $J_{\mathrm{eff}}=1/2,$ split into three separated manifolds. The calculated DOS [see Ref.\ \onlinecite{Comin}, Fig.\ 4(e-f)] clearly demonstrates this feature. Angle-integrated photoemission spectra \cite{Comin} between $-2$ eV and 0 eV were decomposed into four Gaussians, corresponding to the four bands described above, at the positions $-0.5$,$-1$,$-1.4$, and $-1.9$ eV, with weights roughly corresponding (given the possible energy dependence of the matrix elements) to 1:2:1:1, consistent with the band structure calculation \cite{Foyev13}.

Our experimental DOS, which we estimate in Fig.\ 3(b) by averaging the
normalized conductivity, $(dI/dV)/(I/V)$, over the two surfaces, is
largely consistent with this observation.  We note that the higher-bias
DOS is somewhat overestimated in the measurements, probably due to
incomplete cancellation of the tunneling matrix elements and the band edges are distorted due to the calculation process. It is worth
noting that in Ref.\ \onlinecite{Comin}, as well as in our experiment, the
centers of the occupied bands are shifted by 0.1 to 0.3 eV to lower energies which would also increase the minimal gap.

Nevertheless, the decomposition of the angle integrated photoemission spectra between $-2$ eV and $0$ eV corresponding to the four bands described above and the consistence with theory might be fortuitous. Indeed, accurate modeling of the ARPES spectra may require including spectral features of both surface reconstructions (with similar weights), as suggested by our structural analysis.

Comparing our results with Ref.\ \onlinecite{Comin}, the significantly larger band gap found in STS remains an open issue. As ARPES does not directly provide the band gap value, the evolution of ARPES spectra on coating the surface with K was used to determine this value. While Ref.\ \onlinecite{Comin} discussed the deposition of K only as doping,
another reasonable process is the transformation of the $(\sqrt{3}\times
\sqrt{3})\mathrm{R30^{\circ }}$ surface as K is incorporated. Such a global change of the ratio of $1\times 1$ and $(\sqrt{3}\times
\sqrt{3})\mathrm{R30^{\circ }}$ reconstructed areas may also result in a modified ARPES spectrum with a shift of spectral weight to higher binding energies.

Considering the STS spectra, tip-induced band bending might be a possible candidate to explain an apparent increase of the band gap. Our topographic data shows that the surfaces exhibits atomic scale defects leading to the expectation of Fermi level pining inside the band gap similar to results on semiconductor surfaces \cite{FeenstraBB}. This scenario implies that STS does indeed reflect a significantly larger surface band gap.

Finally, we address the issue of QSH behavior.  Ref.\ \onlinecite{Shitade}
predicted that Na$_{2}$IrO$_{3}$ is a QSH insulator and therefore that
the band gap must briefly close as a step is traversed.
However, as we have established, the surface of Na$_{2}$IrO$_{3}$ is
rather different from the bulk. Even if the model of
Ref.\ \onlinecite{Shitade} correctly captures the relevant features of the
bulk band structure, it may be not applicable to the actual surfaces
that are realized in nature.  With this caveat in mind, we have
monitored the tunneling gap as the tip traverses a step (Fig.\
\ref{4}). The observed step heights are in agreement with the crystal
structure \cite{Choi12} and correspond to the height of five unit
cells (26.5 \AA) and two unit cells (10.6 \AA), respectively. Figures
4(c) and (d) show spatially resolved STS spectra taken along the red
lines in Figs. 4(a) and (b). Upon traversing a step separating two
$1\times 1$ surfaces, a smooth transition can be observed with no sign
of gap closure anywhere.  The step separating two $(\sqrt{3}\times
\sqrt{3})\mathrm{R30^{\circ }}$ regions likewise shows a constant band
gap with so sign of closure.  We conclude that neither of the two
possible terminations of Na$_{2}$IrO$_{3}$ shows any evidence of the
QSH effect.

Work supported by the German Science Foundation through SPP 1666 and the Helmholtz Virtual Institute 521. This work was also supported in part by the U.S. Office of Naval Research through the Naval Research Laboratory's Basic Research Program.


\begin{thebibliography}{99}
\bibitem{Kim} B.J. Kim, H. Ohsumi, T. Komesu, S. Sakai, T. Morita, H.
Takagi, T. Arima, Science \textbf{323} 1329 (2009).

\bibitem{Shitade} A. Shitade, H. Katsura, J. Kunesš, X.-L. Qi,  S.-C. Zhang,
N. Nagaosa, Phys. Rev. Lett. \textbf{102}, 256403 (2009).

\bibitem{Pesin} D. Pesin and L. Balents, Nature Phys. \textbf{6} 376 (2010).

\bibitem{Jackeli1} J. Chaloupka, G. Jackeli, and G. Khaliullin Phys. Rev.
Lett. \textbf{105}, 027204 (2010).

\bibitem{Jiang} Hong-Chen Jiang, Zheng-Cheng Gu, Xiao-Liang Qi, S. Trebst,
Phys. Rev. B \textbf{83} 245104 (2011).

\bibitem{Mazin12} I. I. Mazin, H. O. Jeschke, K. Foyevtsova, R. Valent{\'\i}%
, D. I. Khomskii, Phys. Rev. Lett. \textbf{109}, 197201 (2012).

\bibitem{Foyev13} K. Foyevtsova, H. O. Jeschke, I. I. Mazin, D. I. Khomskii,
R. Valent{\'\i}, Phys. Rev. B \textbf{88}, 035107 (2013).

\bibitem{Singh10} Y. Singh and P. Gegenwart, Phys. Rev. B \textbf{82},
064412 (2010).

\bibitem{Feenstra1} R. M. Feenstra, P. M\aa rtensson, Phys. Rev. Lett. 
\textbf{61}, 447-450 (1988)

\bibitem{Ye} C. Ye, P. Cai, R. Yu, X. Zhou, W. Ruan, Q. Liu, C. Jin, Y.
Wang, Nat. Commun. \textbf{4}, 1365 (2013)

\bibitem{kresse_phys_rev_b_1993a}G. Kresse and J. Hafner, Phys. Rev. B
  {\bf 47}, 558 (1993).

\bibitem{Kresse1996}G. Kresse and J. Furthm\"{u}ller, Phys. Rev. B
  {\bf 54}, 11169 (1996).

\bibitem{Tersoff} J. Tersoff, D. R. Hamann, Phys. Rev. B \textbf{31}, 805
(1985)

\bibitem{footnote1}Although it is possible to construct surface models
  having $(\sqrt{3}\times\sqrt{3})\mathrm{R30^{\circ }}$ periodicity
  without fully depleting the topmost Na layer, these are strongly
  contradicted by the observed STM topography.
  
\bibitem{Feenstra2} J. A. Stroscio, R. M. Feenstra and A. P. Fein, Phys.
Rev. Lett. \textbf{57}, 2579-2582 (1986)
  
\bibitem{Comin} R. Comin, G. Levy, B. Ludbrook, Z.-H. Zhu, C.N. Veenstra,
J.A. Rosen, Y. Singh, P. Gegenwart, D. Stricker, J.N. Hancock, D. van der
Marel, I.S. Elfimov, A. Damascelli, Phys. Rev. Lett. \textbf{109}, 266406
(2012).

\bibitem{FeenstraBB}R. M. Feenstra, Y. Dong, M. P. Semtsiv and W. T. Masselink, Nanotechnology \textbf{18}, 044015 (2007) 

\bibitem{Choi12} S. K. Choi, R. Coldea, A. N. Kolmogorov, T. Lancaster, I.
I. Mazin, S. J. Blundell, P. G. Radaelli, Y. Singh, P. Gegenwart, K. R.
Choi, S.-W. Cheong, P. J. Baker, C. Stock, and J. Taylor, Phys. Rev. Lett. 
\textbf{108}, 127204 (2012).



\end{thebibliography}
\end{document}